\documentclass[prl,twocolumn,showpacs,superscriptaddress,floatfix]{revtex4}
\usepackage{graphicx}
\usepackage{dcolumn}

\newcommand{\vin}{\ensuremath{v_{\text{in}}}}
\newcommand{\vdc}{\ensuremath{V_{\text{dc}}}}
\newcommand{\vrf}{\ensuremath{v_{\text{rf}}}}
\newcommand{\lqo}{\ensuremath{q_{0}}}

\newcommand{\ecs}{\ensuremath{E_{c}^{s}}}
\newcommand{\eco}{\ensuremath{E_{c}^{0}}}
\newcommand{\ej}{\ensuremath{E_{J}}}
\newcommand{\rn}{\ensuremath{R_{n}}}
\newcommand{\ngt}{\ensuremath{n_{g}}}
\newcommand{\atot}{\ensuremath{A_{\text{tot}}}}
\newcommand{\gqp}{\ensuremath{\Gamma_{\text{qp}}}}
\newcommand{\gcp}{\ensuremath{\Gamma_{\text{cp}}}}
\newcommand{\dq}{\ensuremath{\delta q}}
\newcommand{\ie}{\textit{i.~e.}}
\newcommand{\eg}{\textit{e.~g.}}
\newcommand{\ec}{\ensuremath{E_{c}}}
\newcommand{\rk}{\ensuremath{R_{K}}}
\newcommand{\ehz}{\ensuremath{e/\sqrt{\mathrm{Hz}}}}
\newcommand{\aehz}[1]{\ensuremath{#1\:\ehz}}
\newcommand{\e}[1]{\ensuremath{\times 10^{#1}}}
\newcommand{\units}[1]{\ensuremath{\mathrm{#1}}}
\newcommand{\amount}[2]{\ensuremath{#1\:\units{#2}}}
\newcommand{\cp}{\ensuremath{C_{p}}} 
\newcommand{\csig}{\ensuremath{C_{\Sigma}}}
\newcommand{\cg}{\ensuremath{C_{g}}}
\newcommand{\tfrac}[2]{\ensuremath{{\textstyle\frac{#1}{#2}}}}
\newcommand{\half}{\tfrac{1}{2}}
\newcommand{\vrfl}{\ensuremath{v_{r}}}
\newcommand{\iv}{$I$-$V$}
\newcommand{\qo}{\ensuremath{Q_{0}}}

\begin{document}

\title{Sensitivity and Linearity of Superconducting Radio-Frequency
Single-Electron Transistors: Effects of Quantum Charge Fluctuations }

\author{Madhu Thalakulam}
\affiliation{Department of Physics and Astronomy, Rice University, Houston, TX 77005 USA}
\affiliation{Rice Quantum Institute, Rice University, Houston, TX 77005 USA}
\author{Z. Ji}
\affiliation{Department of Physics and Astronomy, Rice University, Houston, TX 77005 USA}
\author{A. J. Rimberg}
\affiliation{Department of Physics and Astronomy, Rice University, Houston, TX 77005 USA}
\affiliation{Rice Quantum Institute, Rice University, Houston, TX 77005 USA}
\affiliation{Department of Electrical and Computer Engineering, Rice University,
Houston, TX 77005 USA}

\begin{abstract}
We have investigated the effects of quantum fluctuations of quasiparticles on
the operation of superconducting radio-frequency single-electron transistors
(RF-SETs) for large values of the quasiparticle cotunneling parameter
$\alpha=8E_{J}/E_{c}$, where $E_{J}$ and $E_{c}$ are the Josephson and charging energies. 
We find that for $\alpha>1$, subgap RF-SET operation is still feasible despite
quantum fluctuations that renormalize the SET charging energy and wash out
quasiparticle tunneling thresholds.  Surprisingly, such RF-SETs show linearity
and signal-to-noise ratio superior to those obtained when quantum fluctuations
are weak, while still demonstrating excellent charge sensitivity.
\end{abstract}


\maketitle

The radio-frequency single electron transistor (RF-SET) is a  highly sensitive,
fast electrometer, and has been suggested as a potentially quantum-limited
linear amplifier suitable for measurements of individual electronic charges
\cite{Schoelkopf:1998,Devoret:2000,vandenBrink:2002,Clerk:2002}.  Recent
investigations have addressed use of the RF-SET as an 
electrometer\cite{Schoelkopf:1998,Aassime:2001a,Lehnert:2003a}, a readout device
for charge based qubits \cite{Makhlin:2001a,Aassime:2001,Lehnert:2003}, and a
sensor for real-time electron counting experiments \cite{Lu:2003}.  Linearity is
a fundamental assumption of theoretical discussions of the quantum limits of
amplifiers \cite{Caves:1982,Devoret:2000}. Nonetheless, there has been no
detailed investigation of the range of linear response for the RF-SET\@.

Most theoretical studies of RF-SET performance focus on normal
metal SETs, either in the sequential tunneling
\cite{Korotkov:1999,Devoret:2000,Johansson:2002} or cotunneling regimes
\cite{vandenBrink:2002}, while most experiments are performed using a
superconducting SET (SSET)
\cite{Aassime:2001,Aassime:2001a,Lehnert:2003,Lu:2003}.  Transport in the SSET
can be divided into two regimes, depending on the relative sizes of the bias
voltage \vdc\ and superconducting gap $\Delta$: above-gap ($e\vdc>4\Delta$),
dominated by  Coulomb blockade of quasiparticles, and subgap ($e\vdc<4\Delta$),
dominated by combinations of quasiparticle and resonant Cooper pair tunneling
known as Josephson-quasiparticle (JQP) cycles
\cite{Fulton:1989,vandenBrink:1991a}.  While the best charge sensitivities are
found for above-gap operation \cite{Aassime:2001a}, the SSET backaction---the
rate at which it dephases a measured system---is largest there
\cite{Devoret:2000,Makhlin:2001a,Clerk:2002}.  Recent work has focused on subgap
operation for which backaction is significantly reduced, and shot noise is
non-Poissonian \cite{Choi:2001,Clerk:2002,Lehnert:2003}.  Theoretical studies of
quantum fluctuations in the SSET have been limited to above-gap
cotunneling of quasiparticles \cite{Averin:1997}.  In this Letter we find
that linearity and subgap quantum charge fluctuations in superconducting
RF-SETs are intimately related: as quantum fluctuations
strengthen, linearity and signal-to-noise ratio (SNR) improve, while charge
sensitivity remains excellent.
\begin{figure}
\includegraphics[width=7cm]{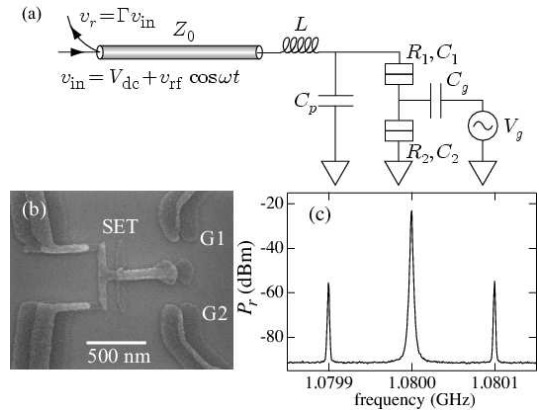}
\caption{\label{sample}  (a) Schematic diagram of the SET illustrating RF
operation.  A voltage \vin\ consisting of dc and RF biases \vdc\ and \vrf\ is
incident on a tank circuit consisting of an inductor $L$, a capacitor \cp, and
the SET, with tunnel junction resistances and capacitances $R_{1(2)}$ and
$C_{1(2)}$. A small charge oscillation $\lqo\cos\omega_{m}t$ modulates the
reflection coefficient $\Gamma$ of the tank circuit and therefore the reflected
voltage \vrfl.  (b) Electron micrograph of S2 (taken after all measurements).
Gates G1 and G2 were used vary the SET offset charge. (c) Power spectrum of
\vrfl\ for $\lqo=\amount{0.063 e}{rms}$ and $\omega_{m}/2\pi=
\amount{100}{kHz}$.  The measured sideband power and noise floor were used to
find the charge sensitivity and SNR of the RF-SET\@.}
\end{figure}

Our SSETs consist of a small island connected to macroscopic leads via two
Al/AlO$_{x}$/Al tunnel junctions J1(2) with normal state resistances $R_{1(2)}$
as illustrated in Fig.~\ref{sample}(a).  We have fabricated and characterized
three samples, S1, S2, and S3 with total resistance $\rn=R_{1}+R_{2}$ of 58, 38
and \amount{24}{k\Omega}; an electron micrograph of S2 is shown in
Fig.~\ref{sample}(b).    The samples were mounted on the mixing chamber of a
dilution refrigerator at its base temperature of \amount{20}{mK}. A Nb chip
inductor $L\approx\amount{120}{nH}$ together with the parasitic  capacitance
$\cp\approx\amount{0.2}{pF}$ of the SET contacts constituted a tank circuit with
resonant frequency $f_{LC}\approx\amount{1}{GHz}$ and quality factor $Q\approx
16$.  We measured the samples' current-voltage (\iv) characteristics in an
asymmetric voltage-biased configuration [Fig.~\ref{sample}(a)] by varying the dc
bias voltage \vdc\ in the absence of an RF excitation.  Details of RF operation
are similar to those discussed elsewhere \cite{Schoelkopf:1998,Aassime:2001a}. 
The SET offset charge $\qo+\lqo\cos\omega_{m}t$ consisted of a dc component \qo\
that set the overall working point and an ac component of amplitude \lqo\ that
modulated the reflected voltage \vrfl. Power spectra of \vrfl\
[Fig.~\ref{sample}(c)] were used to determine the charge sensitivity $\delta q$
and SNR\@.
\begin{table}
\caption{\label{table}Sample parameters.  Resistances are in k$\Omega$, energies
in $\mu$eV, and areas in \amount{10^{-3}}{\mu m^{2}}.}
\begin{ruledtabular}
\begin{tabular}{l|cccccccc}
& \rn & $\Delta$ & \ec & \ej & $\alpha$ & \atot & \eco & \ecs \\ \hline
S1 & 58 & 200 & 230 & 22 & 0.78 & 4.1 & 254 & --- \\
S2 & 38 & 200 & 250 & 34 & 1.08 & 3.4 & 291 & 258 \\
S3 & 24 & 190 & 162 & 54 & 2.65 & 5.0 & 218 & 162 \\
\end{tabular}
\end{ruledtabular}
\end{table}

\begin{figure}
\includegraphics[width=7cm]{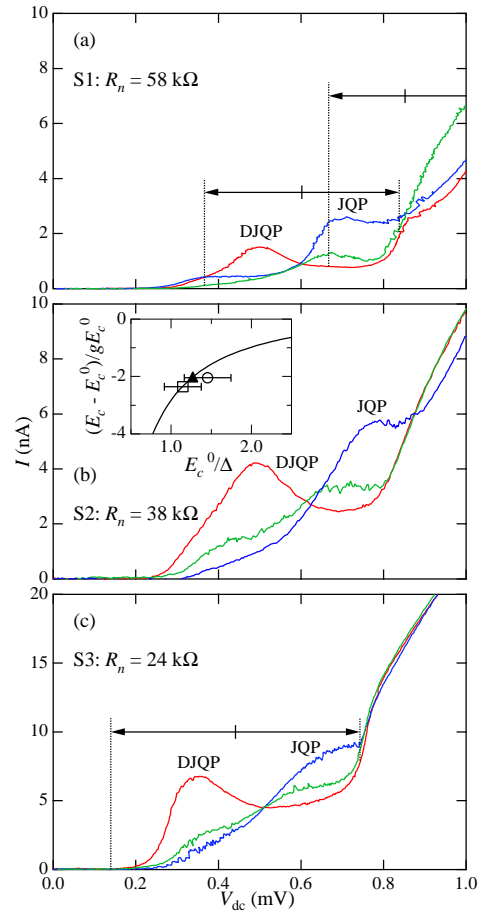} \caption{\label{ivfig} \iv\
characteristics for (a) S1 (b) S2 and (c) S3 (note scale change), were chosen
for \qo\ showing the DJQP process (red), the JQP process (blue), and an
intermediate value of \qo (green).  The arrows and vertical hash mark show the
peak-to-peak RF amplitude $2Q\vrf$ and dc bias \vdc\ for optimal RF-SET
operation.  Inset: variation in the measured charging energy \ec\ relative to
the bare charging energy \eco\ for S1 (solid triangle), S2 (circle) and S3
(square).  Error bars indicate uncertainty in \eco.  Solid line: theoretical
prediction. }
\end{figure}

\begin{figure}
\includegraphics[width=7.0cm]{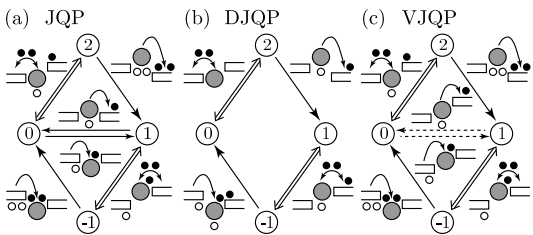}
\caption{\label{cycles} Various JQP cycles.  Here J2(1) is on the
left (right) and $\vdc>0$.  Solid (empty) circles indicate quasielectrons
(quasiholes) created during a cycle. (a) JQP cycle.  Beginning in the state
$n=0$ ($n=1$), where n is the number of excess electrons on the SET, the
transition $0\rightarrow 1$ ($1\rightarrow 0$) is allowed, bringing Josephson
tunneling through J1(2) into resonance.  Cooper pair tunneling 
$1\Leftrightarrow -1$ ($0\Leftrightarrow 2$) is interrupted by quasiparticle
tunneling through the opposite junction $-1\rightarrow 0$ ($2\rightarrow 1$),
completing the cycle.  (b) DJQP cycle.  When Josephson tunneling is
simultaneously resonant through both J1 and J2, transport occurs via the
sequence $0\Leftrightarrow 2$, $2\rightarrow 1$, $1\Leftrightarrow -1$,
$-1\rightarrow 0$.  (c) Proposed VJQP cycle.  If the transition $0\rightarrow 1$
($1\rightarrow 0$) is forbidden, it may still occur virtually.  The remaining
JQP transitions are allowed for relevant \vdc.  }
\end{figure}

In Fig.~\ref{ivfig}, we show representative \iv\ characteristics of the samples
in the superconducting state, measured for different \qo, with $\lqo=0$.  For
S1, we observe clear above-gap ($\vdc \agt \amount{800}{\mu V}$) current
modulation  corresponding to Coulomb blockade of quasiparticle tunneling
[Fig.~\ref{ivfig}(a)].  The main sub-gap features corresponding  to the JQP
\cite{Fulton:1989,vandenBrink:1991a} cycles are sharp and clearly distinguished.
 As illustrated in Fig.~\ref{cycles}, the simplest JQP cycle consists of
resonant tunneling of a Cooper pair through one junction and dissipative
tunneling of two quasiparticles  through the other, transporting two electrons
through the SET\@.  The cycle can occur only when the transition
$0\rightarrow1$ ($1\rightarrow 0$) is allowed, \ie, for $e \vdc>E_{c} + 2\Delta$
where $\ec=e^{2}/2\csig$ is the charging energy of the SET and
$\csig=C_{1}+C_{2}+2\cg$ its total capacitance.  While the JQP cycle is
forbidden at lower bias, at $\qo/e=\ngt\approx\half$ and $e\vdc= 2\ec$ Cooper
pair tunneling is resonant at both junctions and the double JQP (DJQP) cycle
becomes possible.  The fact that sequential tunneling  cannot occur via either
cycle for $2\ec\alt e \vdc \alt \ec +2\Delta$ is reflected in S1 by a sharp drop
in current at $\vdc\approx\amount{630}{\mu V}$ just below the JQP feature.

As \rn\ decreases, so does current modulation for $e\vdc > 4\Delta$, consistent
with suppression of the Coulomb blockade by quasiparticle cotunneling
\cite{Averin:1997}: the modulation is reduced for S2, and nearly absent for
S3 [Fig.~\ref{ivfig}(b) and (c)].  In contrast, features corresponding to the
JQP cycles still exist but become progressively less sharp.  Since these cycles
involve both Cooper pair and quasiparticle tunneling, we hypothesize that subgap
quantum fluctuations of quasiparticles are strong, while quantum fluctuations of
Cooper pairs remain weak. Since to the best of our knowledge no theoretical
description of subgap quantum charge fluctuations in the SSET exists, we provide
simple arguments supporting our hypothesis.

We first compare with known results for above-gap transport. Following
Ref.~\cite{Averin:1997} we define a parameter
$\alpha\equiv\frac{\Delta}{E_c}\frac{\pi\hbar}{e^2}(R_1^{-1}+R_2^{-1})=%
\frac{8\ej}{\ec}$  characterizing the strength of quantum fluctuations for
quasiparticles, assuming $R_{1(2)}=\rn/2$ and using the Ambegaokar-Baratoff
relation for the Josephson coupling energy $\ej=\frac{\Delta}{4}\frac{\rk}{\rn}$
where $\rk=\frac{h}{e^{2}}$.  Quantum fluctuations are negligible for $\alpha\ll
1$.  Determining \ec\ from the location of the DJQP peak and \ej\ from the total
junction resistance we calculate $\alpha$ as in Table~\ref{table}. None of our
samples satisfies $\alpha\ll 1$, although for S1 ($\alpha =0.78)$ some above-gap
Coulomb modulation survives.  The progressively weakening modulation for S2
($\alpha=1.08$) and S3 ($\alpha=2.65$), is consistent with previous results
\cite{Averin:1997}.

Cotunneling as described in Ref.~\cite{Averin:1997} occurs only for
$\vdc>4\Delta/e$: it results in two quasiparticle excitations and transfers a
single electron through the SET.  Other virtual processes, however, remain
important for $\vdc<4\Delta/e$. For normal SETs, \ec\ can be renormalized by
quantum  charge fluctuations: \eg, near $\ngt = 0$, the effective charging
energy $\ec \approx\eco(1- 4g)$ where $g = \rk/\pi^{2}\rn$ is the dimensionless
parallel conductance of the tunnel junctions and \eco\ the bare charging energy;
similar renormalization occurs in the superconducting
state \cite{Joyez:1997,Lehnert:2003a}. Calculating the first-order energy shift
due to transitions $n\rightarrow n\pm 1$, we find the renormalized charging
energy $\ecs = \eco(1 + g
\frac{\Delta}{\eco}{\Gamma[\frac{\Delta}{\eco}(1+2\ngt)]+\Gamma[
\frac{\Delta}{\eco}(1-2\ngt)]})$ where
$\Gamma(x)=\int_{0}^{\infty}K_{1}^{2}(u)e^{-xu}\,du$ and $K_{1}(u)$ is a Bessel
function.

Using the expression for \ecs, we find empirically that $\eco=\amount{254}{\mu
eV}$ gives the measured \ec\ for S1.  We measure the total geometric junction
area \atot\ for the samples with an estimated accuracy of $\pm20\%$, obtaining
the values in Table~\ref{table}.  Setting $\eco = e^{2}/2\csig^{0}$ where
$\csig^{0}=C_{1}^{0}+C_{2}^{0}+2\cg$ and using $2\cg \approx \amount{80}{aF}$, we
obtain $C_{1}^{0}+C_{2}^{0}=\amount{195}{aF}$ as the total unrenormalized
junction capacitance for S1.  Scaling this result according to \atot\ we find
$\csig^{0}$, \eco\ and finally \ecs\ for S2 and S3 [Table~\ref{table}];
agreement is excellent given the uncertainties in \atot. The inset to 
Fig.~\ref{ivfig} shows the relative difference between \ec\ and \eco\ scaled by
$1/g$.  The results agree with theory to within our experimental accuracy,
providing strong evidence that subgap quantum fluctuations of quasiparticles
occur in our samples.

\begin{figure}
\includegraphics[width=7cm]{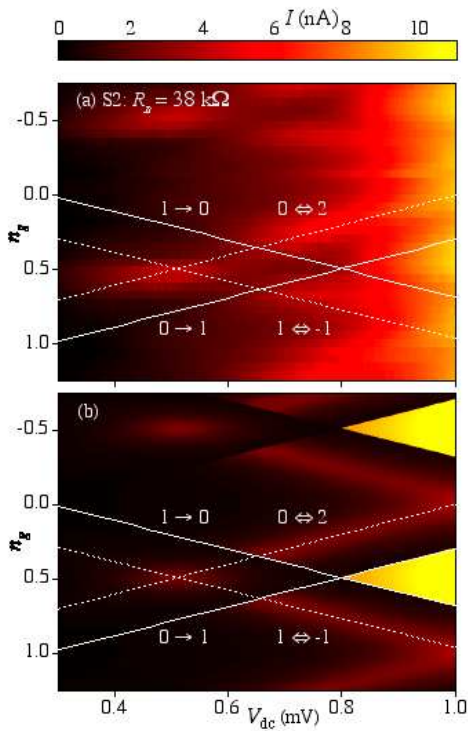}
\caption{\label{image} False color images of $I(\vdc,\ngt)$ for (a) S2 at
$T=\amount{20}{mK}$ (b) a simulation at $T=\amount{200}{mK}$ assuming an
electromagnetic environment with impedance $R_{\text{env}}=\amount{50}{\Omega}$
and temperature $T_{\text{env}}=\amount{1}{K}$.  Cooper pair resonance lines
$0\Leftrightarrow 2$ ($-1\Leftrightarrow 1$) and quasiparticle tunneling
thresholds $1\rightarrow 0$ ($0\rightarrow 1$) are indicated by the dashed and
solid lines.}
\end{figure}

Virtual quasiparticle tunneling may also play a role in subgap transport,
as suggested by the softening of the JQP cycle cutoff in S2 and S3.  To
illustrate such effects more clearly we show a plot of the $I(\vdc,\ngt)$
surface for S2 in Fig.~\ref{image}(a).  The JQP resonances along the
$0\Leftrightarrow 2$ and $1\Leftrightarrow -1$ lines and the DJQP peak at their
intersection are clearly visible, but there is no sharp cutoff of the JQP
process below the $1\rightarrow 0$ ($0\rightarrow 1$) thresholds. For
comparison, in Fig.~\ref{image}(b) we show a simulation of the current in S2
based on sequential tunneling \cite{Lu:2002}  at an elevated temperature and
including photon-assisted tunneling due to an electromagnetic environment. Despite
the extreme conditions the quasiparticle tunneling thresholds are clearly
visible, and the SSET current drops nearly to zero between the JQP and DJQP
features.  The absence of quasiparticle thresholds in Fig.~\ref{image}(a) calls
for an explanation outside of the sequential tunneling picture.

A candidate process that could allow transport along the Cooper pair resonance
lines between the JQP and DJQP features is illustrated schematically in
Fig.~\ref{cycles}(c).  If below threshold the transition $1\rightarrow 0$
($0\rightarrow 1$) occurs virtually, the transitions $0\Leftrightarrow 2$ and
$2\rightarrow 1$ ($-1\Leftrightarrow 1$ and $1\rightarrow 0$) are allowed,
completing what we call the virtual JQP (VJQP) cycle.  Two quasiparticle
excitations are created, but two electrons are transferred through the SET, so
that the process should be allowed for $eV>2\Delta$.  The energy barrier $E_{b}$
for $1\rightarrow 0$ ($0\rightarrow 1$) vanishes at threshold and climbs to
$E_{b}\approx\ec+2\Delta$ at the DJQP peak.  The process can be neglected if the
allowed quasiparticle tunneling rate \gqp\ is small compared to the inverse
dwell time of the virtual quasiparticle: $\gqp\ll E_{b}/\hbar$. Using $\gqp =
4\Delta/e^{2}\rn$, this becomes $\rn\gg\frac{\rk}{\pi}\frac{2\Delta}{E_{b}}$,
which is violated for a range of voltages between the DJQP and JQP features.  A
detailed theoretical analysis is required to determine the contribution of  the VJQP cycle
to transport.

In contrast to the quasiparticle thresholds, features associated with Cooper
pair tunneling are visible in both the data and the simulation, suggesting that
the number of Cooper pairs is well defined.  For the JQP process at resonance,
the Cooper pair tunneling rate is 
$\gcp\approx\ej^{2}/\hbar\gqp=\frac{\pi}{8}\frac{\ej}{\hbar}$
\cite{Averin:1989a}. Demanding that energy broadening due to Cooper pair
tunneling be small compared to the typical energy barrier $4\ec$ for virtual
tunneling gives $2\hbar\gcp/4\ec=\frac{\pi}{16}\frac{\ej}{\ec}\ll 1$, which is
easily satisfied even for S3. For S2 and S3, then,  quantum fluctuations are
significant for quasiparticles but small for Cooper pairs.
 
We now turn to RF operation.  Optimal operating conditions were selected as
follows: a small charge oscillation $\lqo \approx 0.006 e$~{rms} was applied and the
SNR determined from the power spectrum of \vrfl\ as in Fig.~\ref{sample}(c). 
Subgap operation (all samples) and above-gap operation (S1) were optimized
over dc bias \vdc, rf bias \vrf\ and offset charge \qo.  We measured SNR versus
input amplitude \lqo\ for each optimization and determined the charge
sensitivity $\delta q$ using  $\delta
q=\frac{\lqo}{\sqrt{\text{BW}}}10^{-\text{SNR}/20}$ where the resolution
bandwidth $\text{BW}=\amount{1}{kHz}$ and SNR is in dB \cite{Aassime:2001a}.

\begin{figure}
\includegraphics[width=7cm]{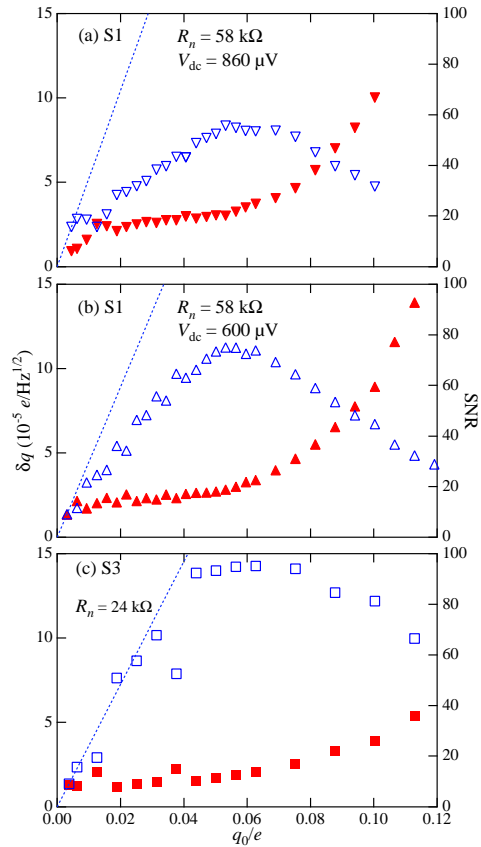}
\caption{\label{cssnr} Charge sensitivity $\delta q$ and SNR (linear scale)
versus \lqo\ in $e$~rms for (a) S1, above gap, (b) S1, subgap, and (c) S3,
subgap.  Charge sensitivity (solid red symbols) is plotted on the left axis and
SNR (open blue symbols) on the right. For reference, the SNR for linear response
is plotted as the dashed lines for \dq\ measured at the
smallest \lqo. }
\end{figure}

The optimized biases for S1 and S3 are indicated in Fig.~\ref{ivfig} and the
results of the $\delta q$ and SNR measurements in Fig.~\ref{cssnr}.   For S1
the best $\dq=\aehz{9\e{-6}}$ was found for
$\vdc=\amount{860}{\mu V}$, consistent with previous results
\cite{Aassime:2001a}.  Linearity, however, was poor: as \lqo\ increases, the
measured SNR rapidly becomes sublinear, and \dq\ worsens [Fig.~\ref{cssnr}(a)]. 
Since \dq\ apparently does not saturate even for $\lqo=\amount{4.5\e{-3}e}{rms}$
it is unclear how small \lqo\ must be to achieve linear response. For subgap
operation ($\vdc=\amount{600}{\mu V}$) of S1 [Fig.~\ref{cssnr}(b)], we find $\dq
\approx \aehz{1.3\e{-5}}$, with SNR nearly linear to $\lqo\alt\amount{0.01
e}{rms}$. Since \dq\ appears close to saturation at $\lqo=\amount{3.1\e{-3
}e}{rms}$, we may have approached linear response.
  
For S3 the best operating point occurred at
$\vdc=\amount{440}{\mu V}$ [Fig.~\ref{cssnr}(c)], between the DJQP and JPQ features with
$\dq\approx\aehz{1.2\e{-5}}$, better than that for subgap operation of S1.
Moreover, linearity was vastly improved: the SNR remains linear and \dq\
nearly flat to $\lqo=\amount{0.038 e}{rms}$ indicating that we have achieved
linear response in this sample.  For S2 (data not shown) the
best $\dq\approx\aehz{1.2\e{-5}}$ also occurred subgap, and the SNR was linear
to $\lqo\approx\amount{0.02 e}{rms}$.

We can now make some general statements about the effects of quantum
fluctuations on RF-SET operation.  For samples with smaller $\alpha$ such as S1,
transport is fairly well described by the sequential tunneling picture: \iv\
characteristics are sharp and vary strongly with \qo\, giving rise to excellent
charge sensitivity.   The same sharpness, however, prevents good linearity,
since a large \lqo\ necessarily moves the SET far from  optimal operation.  For
samples with larger $\alpha$ such as S3, quantum fluctuations cause at least two
important effects.  First, the subgap features are smoothed and broadened,
improving linearity: \eg, in S3 there is no ``dead spot'' between the DJQP and
JQP features for which the SSET current is roughly independent of \qo.  Second,
renormalization of \ec\ moves the DJQP feature to lower bias, so that the
optimal rf amplitude of about $(2\Delta - \ec)/e$ increases with $\alpha$.
Finally, the smaller \rn\ simplifies impedance matching between the RF-SET and
the \amount{50}{\Omega} coaxial line.

In conclusion, we have investigated the influence of quantum charge fluctuations
on the charge sensitivity and SNR of RF-SETs. We find that RF-SETs with
$\alpha\agt 1$--2 (strong quantum fluctuations) show both good linearity and
good charge sensitivity.  In contrast, RF-SETs with $\alpha<1$ (weak quantum
fluctuations) show poor linearity and only modestly better charge sensitivity. 
These findings assume particular importance given interest in  the RF-SET as a
potentially quantum-limited linear amplifier.  We have achieved linear response
only for subgap operation in samples with $\alpha\agt 1$ for which quantum
fluctuations of quasiparticles are substantial.

This work was supported by the National Science Foundation under Grant No.\
DMR-0242907, by the Army Research Office under Contact No.\ DAAD19-01-1-0616 and
by the Robert A. Welch Foundation.  We thank P. D. Dresselhaus and S. P. Benz for fabrication of the Nb chip inductors and S. M. Girvin for a critical reading of the manuscript.


\end{document}